\documentclass[12pt]{iopart}

\usepackage{graphicx}
\usepackage{ulem}
\usepackage{color}
\usepackage{colortbl}
\usepackage{subfig}

\usepackage{dcolumn}
\usepackage{bm}
\usepackage[titletoc,title]{appendix}
\usepackage[utf8]{inputenc}
\usepackage{comment}

\newcommand{\beq}{\begin{eqnarray}}
\newcommand{\eeq}{\end{eqnarray}}
\newcommand{\bit}{\begin{itemize}}
\newcommand{\eit}{\end{itemize}}

\newcommand{\bvecr}{{\bf r}}
\newcommand{\bp}{{\bf p }}

\newcommand\z{\zeta}

\newcommand\ta{\theta}

\newcommand\ph{\varphi}

\newcommand\eps{\epsilon}

\newcommand\di{\textrm{d}}
\newcommand\ex{\textrm{e}}

\begin{document}

\title[Physical phase field model for phagocytosis]{Physical phase field model for phagocytosis}

\author{Benjamin Winkler$^{1}$, Mohammad Abu Hamed$^{2,3,4}$, 
Alexander A. Nepomnyashchy$^{2}$, Falko Ziebert$^{5}$ }

\address{$^1$Physikalisch-Technische Bundesanstalt, 10587 Berlin, Germany}
\address{$^2$Department of Mathematics, Technion - Israel Institute of Technology, 32000 Haifa, Israel}
\address{$^3$Department of Mathematics, The College of Sakhnin - Academic College for Teacher Education, 30810 Sakhnin, Israel}
\address{$^4$Department of Mathematics, Gordon College of Education, 3570503 Haifa, Israel}
\address{$^5$Institute for Theoretical Physics, Heidelberg University, 69120 Heidelberg, Germany}
\ead{f.ziebert@thphys.uni-heidelberg.de}
\vspace{10pt}

\begin{abstract}
We propose and study a simple, physical model for phagocytosis, i.e.~the active, 
actin-mediated uptake of micron-sized particles by biological cells. 
The cell is described by the phase field method
and the driving mechanisms of uptake are actin ratcheting, modeled by a dynamic vector field, 
as well as cell-particle adhesion due to receptor-ligand binding.  
We first test the modeling framework for the symmetric situation of a spherical cell
engulfing a fixed spherical particle. 
We then exemplify its versatility by studying various asymmetric situations 
like different particle shapes and orientations, 
as well as the simultaneous uptake of two particles. 
In addition, we perform a perturbation theory of a slightly modified model version  
in the symmetric setting, allowing to derive a  reduced model, 
shedding light on the effective driving forces and being easier to solve.
This work is meant as a first step in describing phagocytosis and we discuss
several effects that are amenable to future modeling within the same framework.   
\end{abstract}

%
%
%
%
%

\section{Introduction}

Biological cells often have to engulf larger particles \cite{portals_of_entryNature}, 
either because they feed on them or as part of the immune response
against intruders like bacteria, or dying cells.
While the cell membrane's primary function is that of a barrier, 
it has to be much more versatile and dynamic  \cite{Felix_reviewSM}
and hence cells have developed several mechanisms to ingest 
 particles:
uptake of small (up to $\sim 100$ nm) particles is called endocytosis and is typically passive,
mediated by receptor-ligand binding or clathrin self-assembly \cite{FreySM}.
In contrast, larger micron-sized particles are  {\it actively} taken up
by a process called  phagocytosis. Finally,  
there is a third and less specific way, called  macropinocytosis, 
where membrane protrusions are actively formed
``out of the blue'' and the cell ``swallows'' fluid that may contain several particles
\cite{Shaping_cups_review}. 

Phagocytosis modeling has been reviewed recently in  Ref.~\cite{EndresRPP}. 
Models mostly focus on the so-called zipper mechanism \cite{endres_zipper}
of receptor-ligand binding, overcoming membrane restoring forces.
In fact, in most phagocytes, Fc receptors (FcR) in the cell membrane are
attaching to immunoglobulin G (IgG) present on the particle's surface, 
leading to a zipper-like advancement of a thin membrane protrusion
around the particle. Refinements of such models also account
for signaling and coupling to active processes, phenomenologically modeling the effects of actin.
On the other hand, a framework focusing more on flows and forces, and 
employing a two-fluid model, for the cytoplasm and the actin network, 
was developed in Refs.~\cite{Herant2005,Herant2006}.
Very recently, a Monte-Carlo approach for the membrane shape, 
including ``active energy'' from actin,
was proposed as well \cite{Govprep}.

While it was clear early on that self-assembling, biochemically driven actin filaments play
some role  \cite{phago_recptor_to_actin}, 
recent experiments showed more explicitly than ever 
that particle internalization in phagocytosis is driven by the formation
of 
actin protrusions that wrap around the particle \cite{Waterman}.
It is therefore crucial to develop models including the actin dynamics 
underneath the deformable and dynamic membrane close to the particle. 
Upon closer inspection, it turns out that phagocytosis
and cell motility are quite related to each other:
in fact, both rely on actin-driven membrane protrusions, on the adhesion of these
to a substrate (round particle vs.~often flat substrate, FcR-IgG vs.~integrin binding to fibronectin 
or related ligands), and confront us with the numerical problem of moving and deformable boundaries.
Luckily, the modeling of cell motility has seen much progress in recent years,
mostly due to the phase field framework \cite{FZ_review}.
Just recently, a phase field approach  has  already been applied to
macropinocytosis \cite{phasefield_macropino}, however implementing 
activator-inhibitor fields instead of actin.

We here propose a simple, physical modeling framework to model the initial stages of phagocytosis.
It is based on a three-dimensional (3D) phase field model for crawling cells \cite{winkler_cell3D}, 
only slightly adapted to the problem under consideration. 
We show the framework's versatility by studying the uptake dynamics
of spheres as a function of actin ratcheting, 
the effects of particle shape and orientation, 
and the simultaneous uptake of two particles. 
In addition, we use perturbation theory to derive a reduced equation
for the cell shape, valid for the initial regime of phagocytosis,
that is easy to interpret and faster to solve.

\section{Basic model framework}
\label{sec1}

We base ourselves on a phase field modeling framework that was 
originally developed for cell motility, first in 2D 
\cite{ziebert2011model} and then in 3D \cite{winkler_cell3D}.
We stick here to the basic version of this framework, to make
the model as simple and transparent as possible. 
Importantly, however, many model extensions and variations have already been included 
in the context of cell motility, for instance,
adhesion formation to inhomogeneous substrates \cite{lober2014modeling},
flows \cite{ShaoPNAS} and elastic effects \cite{RobertEPJE} in the cytoskeleton,
as well as regulation pathways \cite{voigtcell} and activator-inhibition dynamics \cite{Alonso},
which all pave routes to future refined models for phagocytosis as well.

The approach is conceptually simple and flexible and has two dynamic variables: 
the cell's shape is described by a dynamic 3D scalar phase field $\rho(\bvecr,t)\in[0,1]$
and its transition region is identified with the cell membrane
($\rho=1$ corresponding to the cell
and $\rho=0$ to the outside).
The second dynamic variable is a 3D vector field $\bp(\bvecr,t)$ that ``lives'' only inside the domain
defined by $\rho$ and that describes 
the mean degree of parallel ordering (absolute value)
and the mean orientation (vectorial direction)
of the structurally and dynamically polar actin filaments in the cytosol.
Additional phase field(s), that can be either static or dynamic, 
can then be used to define regions with steric exclusion, 
allowing to implement arbitrary substrates or environments \cite{winkler_cell3D}.
In the case of interest here, we will use such an additional phase field
$\Phi(\bvecr)$ to describe a particle of a chosen shape 
that may get close to the cell and internalized via phagocytosis.

The actin dynamics inside the cell is implemented on the phenomenological level
via its basic physical features: 
actin is nucleated close to the membrane (i.e.~the phase field interface)
and exerts a force against the membrane via the well-known 
polymerization ratchet mechanism \cite{Mogilner_number},
pushing  it forwards.
Two more properties of actin are important:
i) its nucleation is localized also close to the substrate/particle since its regulators 
receive signals there.  This is implemented via another phase field
$\Psi(\mathbf{r})$ (which, if not stated otherwise, we choose here to be the same as $\Phi(\bvecr)$ for simplicity) 
having a certain decay length away from the substrate/particle\footnote{
Both $\Phi(\mathbf{r})$ and $\Psi(\mathbf{r})$ are, in the simplest case of a spherical bead, 
implemented as a function with a tanh-like radial profile
around the particle's center, see also  \ref{numerics} for more information
about the implementation.
}
and as such defining a region where actin nucleation is possible \cite{CatesNCom}.  
And ii), actin is oriented predominantly tangentially to the local substrate/particle surface,
which can be modeled via a projection operator onto the local tangential plane. 

Explictly, the 3D model equations read
\beq\label{eq1}
\partial_t \rho &=& D_{\rho}\Delta \rho - \rho(1-\rho)(\delta[\rho]- \rho)- \alpha \mathbf{p} \cdot \nabla\rho - \kappa \nabla\Phi\cdot\nabla\rho - \lambda \rho\,\Phi^2\,,\\
\label{eq2}
\partial_t\mathbf{p} &=& 
D_{p}\Delta\mathbf{p} 
- \beta \Psi \hat{P}(\nabla\rho) 
 -\tau^{-1}\mathbf{p}-\Phi^2\mathbf{p}\,.
\eeq 
with the two static fields $\Phi(\bvecr), \Psi(\bvecr)$ defining the particle, 
and the localization of its stimulation of actin polymerization 
close to that particle, respectively. 

The first two terms in Eq.~(\ref{eq1}) describe  classical phase field dynamics
\cite{ziebert2011model,FZ_review}:
$D_\rho$ fixes the width of the diffuse interface 
and the second term implements a relaxational dynamics in the phase field 
double well potential.
Since in this simplest implementation the dynamics of $\rho$ is non-conserved, 
it needs to be supplemented by a volume conservation, 
implemented here via a global constraint, 
$
\delta[\rho]=\frac{1}{2}+\mu\left[\int\hspace{-1mm}\rho\,d^3\mathbf{r} - V_0\right],
$ 
with the second term penalizing differences
between the cell's current volume $\int\hspace{-1mm}\rho\,d^3\mathbf{r}$
and the prescribed volume $V_0$,
the stronger the larger the ``stiffness'' $\mu$.

The last three terms in Eq.~(\ref{eq1}) model processes that can 
move the phase field boundary:
$\alpha\bp \cdot\nabla\rho$ implements the pushing by actin ratcheting 
against the cell's boundary, with $\alpha$ an effective velocity
of the interface movement \cite{ziebert2011model}.
$\kappa \nabla\Phi\cdot\nabla\rho$ implements adhesion 
\cite{nononumura2012study,JakobSciRep} between the cell and the particle described
by $\Phi(\mathbf{r})$ by means of ligand-receptor bond formation,
with $\kappa$ the adhesion strength parameter.
Finally, the last term models excluded volume interaction
with the particle; it can be derived from an interaction
potential $\frac{\lambda}{2}\int\rho^2\Phi^2\,d^3{\mathbf{r}}$
with excluded volume strength $\lambda$.
For more information on these contributions see Ref.~\cite{ebookchap1}.

The actin dynamics, Eq.~(\ref{eq2}), has three contributions
from actin turnover:
a diffusion/elastic term, 
a source term  proportional to the
polymerization rate $\beta$  
and  a sink term proportional to $\tau^{-1}$ 
describing depolymerization,
with $\tau$ of the order of the actin turnover time.
Since the actin creation and its pushing via the polymerization ratchet are closely related, 
we typically fix $\beta=3\alpha$,
which was a reasonable value when applying the model 
in the context of cell motility \cite{winkler_cell3D}.  
The geometric structure of the actin source term was developed in Ref.~\cite{winkler_cell3D}:
actin polymerization is modeled anisotropically with respect to the 
local tangential plane defined by $\Phi(\mathbf{r})$
modeling the particle. This is achieved by
employing the projection operator onto that plane, 
$\hat{P} = \hat{I} - \hat{\mathbf{n}}\otimes\hat{\mathbf{n}}$
with $\hat{I}$ the identity and $\hat{\mathbf{n}}=\nabla\Phi/|\nabla\Phi|$ 
the local normal vector on the particle's surface.
Finally, the last term in Eq.~(\ref{eq2}) is
added to ensure that there is no actin penetrating into the particle,
which would be unphysical.

It should be noted that the used phase field approach has an inherent 
surface tension associated with the phase field's wall energy \cite{FZ_review}.
This apparent surface tension is related to the excess energy due to the 
phase field interface and is $\Sigma\propto\sqrt{D_\rho}$.
The surface tension could be removed \cite{Folch1} or tuned
to a desired value \cite{benjPhysD}, but we keep it here,
since it reflects the only restoring forces due to membrane shape changes
(since we do not consider membrane bending yet),
which is a relevant effect in phagocytosis.

\begin{figure}[t!]
\includegraphics[width=0.99\linewidth]{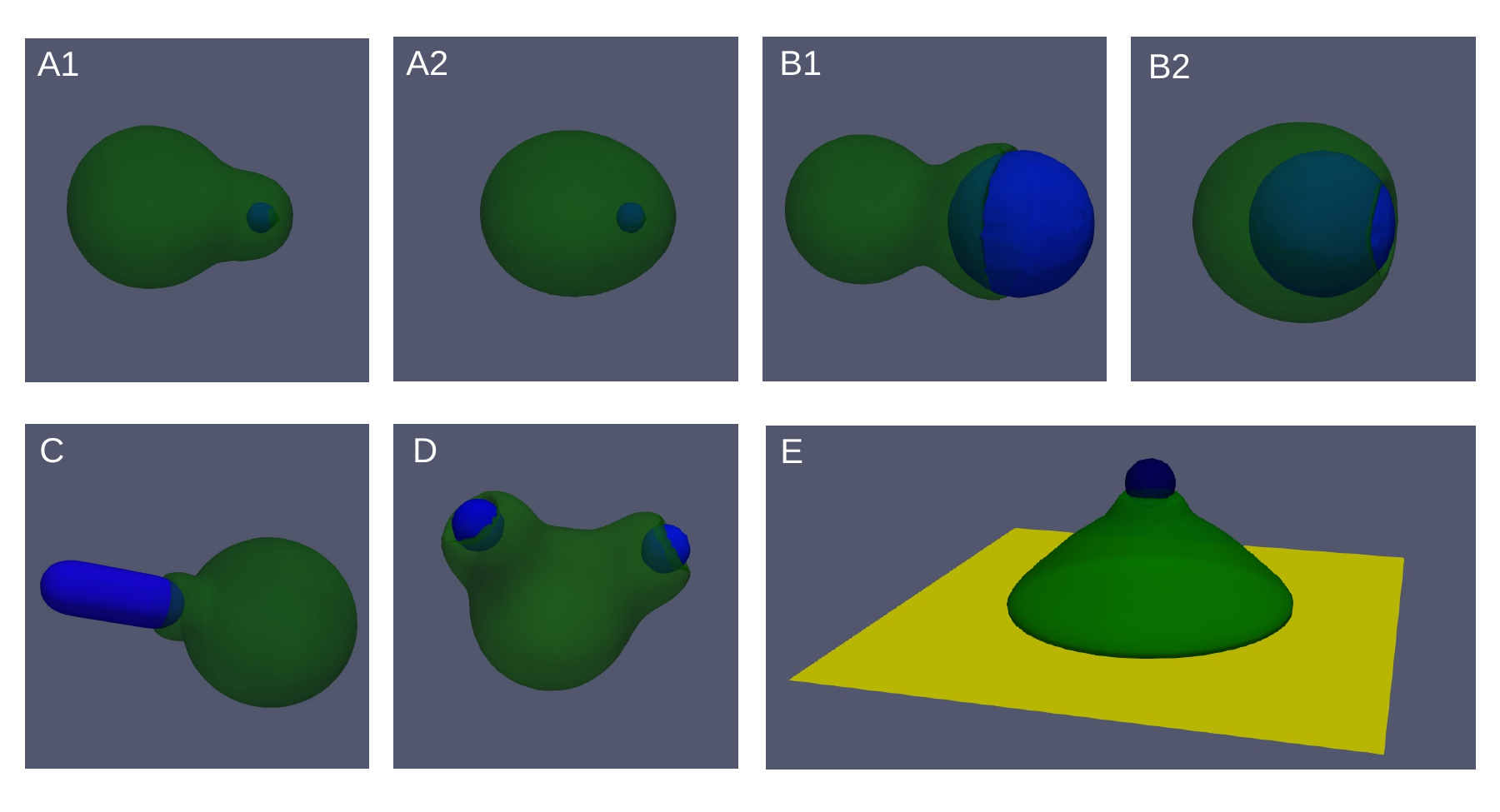}
\caption{
\label{fig_teaser}
Several phagocytosis scenarios that can be treated within the modeling framework.
A1, A2) Early and late stage of a cell ($r_c=14$) engulfing a  small spherical particle ($r_p=3$).
B1, B2) A cell engulfing a particle of size similar to its own ($r_p=11$ vs.~$r_c=14$).
Here, the first phase with the formation of a phagocytic cup is clearly visible, 
followed by a slower relaxation, where the whole cell body 
rounds up. 
Snapshots correspond to $t=50s$ (A1, B1) and $t=400s$ (A2, B2). 
C) A spherocylinder (radius $r=4$;  cylindrical element of length $4r$) 
is engulfed in a "tip-first" configuration; $t=15s$. 
Sufficiently high actin activity 
leads to full engulfment after about $t=400s$.
D) Simultaneous engulfment of two particles (of size $r_p=4$),
positioned at an angle of $90^\circ$ with respect to the initial center of the cell; 
early stage at $t=10s$.
E) A cell was allowed to spread on a flat substrate (yellow) modeled by an additional static phase field. 
Then a small sphere was approached.
Parameters: A, B, D: $\alpha=4$; C: $\alpha=6$; others as given in table \ref{param_table}.
E: as specified in the main text, see section \ref{asym_problems}.
}
\end{figure}

Details on the numerical implementation of the model can be found
in \ref{numerics} with typical parameter values given in table \ref{param_table}. 
Fig.~\ref{fig_teaser} shows applications
to several example geometries relevant for phagocytosis, 
highlighting the versatility of the modeling framework: 
A1 and A2 show the engulfment of a spherical particle of small size,
while in B1 and B2 the particle is of similar size as the phagocytosing cell.
C shows the initial phase of the uptake of a spherocylindrical particle.
D exemplifies the simultaneous uptake of two small spheres and E
the uptake of a sphere by a cell that is  spreading on a flat substrate.

\section{Results for a spherical cell engulfing a spherical particle}
\label{sphere_sphere}

We first applied the model to the most symmetric and most studied situation: 
an initially perfectly round, freely floating cell 
that engulfs a static, non-movable, round bead, cf. Fig.~\ref{fig_teaser} A,B.
This situation has been experimentally studied, e.g., in Refs.~\cite{Herant2005,Herant2006} 
using a doube-micropipette apparatus 
and is often used as a benchmark for more complex geometries and scenarios \cite{EndresRPP}.

To quantify the uptake dynamics, we introduce the following observables:
first, the relative cell-particle contact area, $A(t)\in[0,1]$, which we define 
in the phase-field sense as the integrated overlap
of the cell phase field, $\rho$, and the static phase field describing the particle, $\Phi$:
\begin{equation}
A(t)=\frac{1}{A_{full}}\int \Phi(\mathbf{r})\rho(\mathbf{r},t)d^3\mathbf{r}.
\end{equation}
Here $A_{full}$ is the overlap for full engulfment, obtained by a reference simulation
placing a particle in the center of the cell for the same parameters.
Clearly, $A\in[0,1]$ holds, with $A=1$ meaning full engulfment.

To also quantify the mechanism of phagocytosis, 
as a second observable we measure the total amount of actin generated in the cell, 
normalized by the total cell volume:
\begin{equation}
P(t)=\frac{1}{\int \rho(\mathbf{r},t)d^3\mathbf{r}}\int |\mathbf{p}(\mathbf{r},t)|d^3\mathbf{r}.
\end{equation}
Note that actin is only nucleated in a region close to the particle\footnote{In the case of a cell that in addition spreads by actin-polymerisation, cf.~Fig.~\ref{fig_teaser}E, 
one would need to generalize the definition of $P(t)$ accordingly.}, 
defined by the field
$\Psi(\mathbf{r})$, hence $P(t)$ directly quantifies the amount of actin in the phagocytic cup.\\

\begin{figure}[t!]
\centering
\includegraphics[width=0.99\linewidth]{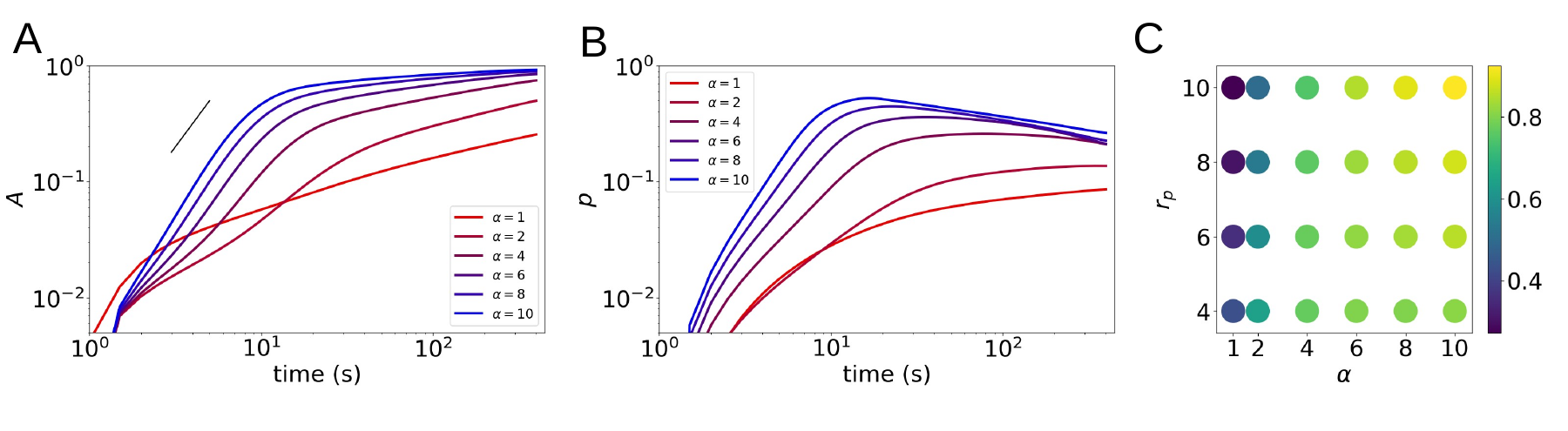}
\caption{
\label{fig_uptakequant}
Quantification of the uptake dynamics of spherical beads
by initially spherical cells. 
A) shows the relative cell-particle contact area, $A(t)$,
for $r_c=14$ and  $r_p=10$ and for different actin pushing rates $\alpha$.
For sufficiently large $\alpha$, there is a fast initial dynamics 
(black line indicates $t^2$), followed by a second slower regime.
B) shows the  total amount of actin, $P(t)$, for the same situation as in A.
C) Shows the final engulfment  $A(t\rightarrow\infty)$ color-coded 
as a function of actin strength and particle radius.
With increasing actin polymerization and ratcheting parameter $\alpha$, 
particle uptake becomes faster and more complete. 
The size dependence is weak, since larger particles also stimulate more actin growth.
Parameters as given in table \ref{param_table}.
}
\end{figure}

\noindent  {\bf Uptake dynamics.} 
Fig.~\ref{fig_uptakequant}A shows the relative cell-particle contact area as a function of time, $A(t)$,
for a cell of radius $r_c=14$ engulfing a large particle of $r_p=10$
(cf.~Fig.~\ref{fig_teaser} panels B1, B2 for two snapshots) 
and varying actin pushing rate $\alpha$. Note that
the actin polymerization rate then also increases  since we chose $\beta=3\alpha$.
For sufficiently large  $\alpha$, one can see an initial fast dynamics 
(the black line shows the power law $A(t)\propto t^2$ as a guide to the eye) 
associated with the cup phase.
This is followed by a second regime of much slower engulfment dynamics.
For smaller $\alpha$ the uptake is slower overall, and the two regimes are less distinct.

Engulfment dynamics has been carefully analyzed in Ref.~\cite{Endres_BP14},
using data from Ref.~\cite{Herant2006},
and interpreted using a receptor-based model. 
There it was found that the engulfed arc length
scales like $\sqrt{t}$ for diffusive receptors and linearly in $t$ for perfect receptor drift.
Our model reflects the second case -- note that we measure the area of contact, $A(t)\propto t^2$, 
which scales like arc length squared  -- due to the constant pushing of actin. 
Experimentally, the behavior is much more complex,
with an apparant jump between a diffusive and an imperfect drift regime,
that could be modeled only by using a signaling molecule in addition \cite{Endres_BP14}.

Fig.~\ref{fig_uptakequant}B shows the volume-normalized total amount of actin, $P(t)$,
that is generated in the cell due to the interaction with the particle. 
Note that only in a region close to the particle (modeled via the static phase field $\Psi$) 
actin generation is stimulated. Its dynamics is characterized by the actin-related 
parameters $\alpha$ 
(reflecting ratcheting) and $\beta$ (nucleation). 
After the initial cup phase, the actin forms a ring-like structure at the protrusion front 
that moves forward and engulfs the particle.
For sufficiently large $\alpha$, $P(t)$ hence increases in time, reaches a maximum at 
roughly half engulfment, and then decreases again. 
This behavior is absent for small $\alpha$ values.
The maximum in $P(t)$ is mostly a geometric effect, since the actin ring 
forming around the particle at the advancing front
is largest at half-wrapping.

Fig.~\ref{fig_uptakequant}C shows an ``uptake diagram'' as a function of
actin ratcheting strength $\alpha$ and particle radius $r_p$ (for fixed cell radius $r_c=14$). 
The color code marks the ``final engulfment'', $A(t\rightarrow\infty)$,
obtained by running long simulations and extrapolating. 
It is clearly visible that increasing $\alpha$ fosters engulfment,
cf.~also supplementary movies 1 \& 2. 
The particle size dependence is rather weak,
with larger particles being engulfed slightly easier.
This is due to the fact that the particle stimulates the generation of actin,
hence a larger bead directly translates to more actin (cf.~also Fig.~\ref{fig_uptakequant}B).
The simple model proposed here so far neglects effects like limited resources for actin 
or time-delay due to actin having to diffuse into the engulfing zone,
which could be included in a future, more refined model. 
 
We also investigated the effect of surface tension.
As discussed  in  section \ref{sec1},
the phase field method has an inherent surface tension 
associated with the phase field's wall energy.
It can be suppressed by using an additional term like $+D_\rho\,c|\nabla\rho|$, with 
$c=-\nabla\cdot \left(\nabla\rho/|\nabla\rho|\right) $ 
the curvature of the cell's phase field interface, 
in Eq.~(\ref{eq1}) \cite{Folch1,benjPhysD}.
Doing so, we generally observe an enhanced engulfment capability of the cell (data not shown). 
The effect is most pronounced in an intermediate parameter regime, 
where shape changes are already significant but the active forces 
are not large enough to quickly and fully engulf the particle.
This confirms that, as expected, surface tension comprises a restoring force 
to interface deformations and hence counteracts phagocytosis.

\section{Select asymmetric problems}
\label{asym_problems}

\begin{figure}[t!]
\includegraphics[width=0.99\linewidth]{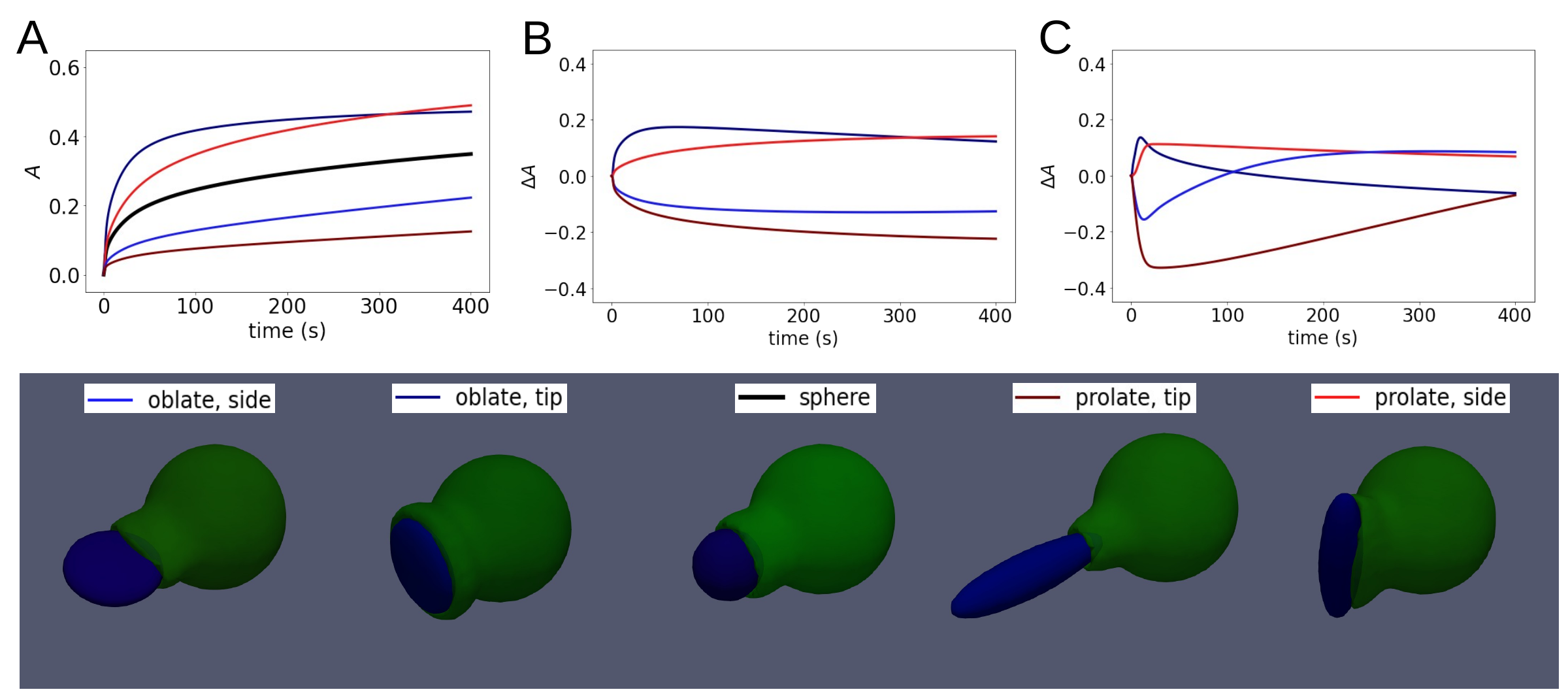}
\caption{
\label{fig_uptake_ellips}
Effect of particle shape on uptake dynamics.
A) shows the relative cell-particle contact area, $A(t)$, as a function of time
for a sphere (black), for an oblate ellipsoid touching the cell with its bottom (violet) 
and with its rim (blue), and for a prolate ellipsoid touching the cell with its side (red)
or tip (brown). All particles have equal volume.
Parameters are such that there is strong receptor-ligand
mediated adhesion ($\kappa=12$) but weak actin ratcheting ($\alpha=1$).
B) Same as A, but now the respective $A(t)$ of the sphere is substracted.
C) Same as B, but now for larger actin ratcheting ($\alpha=4$).
Bottom: Snapshots for the respective shapes and orientations. 
Shown is the case with $\alpha=4$ at $t=10$.
}
\end{figure}

{\bf Effect of particle shape.} Phagocytosis is known to depend on particle shape \cite{ChampionPNAS,endres_shapedepend,phago_elongated}.
While spherocylinders, cf.~Fig.~\ref{fig_teaser}C, are relevant 
since bacteria often have these shapes, 
the most studied and instructive shape variations
are rotational ellipsoids (spheroids).
Experimentally, in Refs.~\cite{SharmaSmith,Paul} it was found that
oblates are engulfed easier than spheres, which in turn are easier to engulf
than prolates.  
In addition, for non-spherical particles the orientation in which 
it meets the cell also is important.
In \cite{ChampionPNAS} the simple idea was put forward
that the curvature at the initial point of contact determines 
-- at least the initial phase of -- phagocytosis. 

Fig.~\ref{fig_uptake_ellips} compares the uptake dynamics of a sphere,
an oblate ellipsoid (``lentil'') and a prolate ellipsoid (``rugby-ball''),
with the ellipsoids in the two possible main orientations.
The volume of all particles was chosen equal, and the aspect ratios of the elliposids
was chosen to be 1/4 and 4, respectively. 
In Fig.~\ref{fig_uptake_ellips}A we consider the case where
receptor-ligand mediated adhesion ($\kappa=12$) dominates over 
actin ratcheting ($\alpha=1$). One can see 
from the dynamics of the relative cell-particle contact area $A(t)$
that less curvature is beneficial
for the initial (so-called pedestal) phase. 
The uptake processes are in the following order:
oblate (bottom) has the lowest curvature and is fastest,
followed by the prolate (side), the sphere, the oblate (rim) and finally the prolate (tip)
which has highest curvature and is slowest.
This is in accordance with receptor-based models, which show 
that higher-curved regions are harder to engulf \cite{EndresRPP}.
For better comparison, Fig.~\ref{fig_uptake_ellips}B
also shows the deviations from the $A(t)$ of the sphere. 

Fig.~\ref{fig_uptake_ellips}C shows the
deviations from the $A(t)$ of the sphere 
for the case where actin ratcheting is more important ($\alpha=4$). 
Interestingly, the initial phase has the same ordering, but then
the least curved oblate (bottom) slows down substantially and the prolate (side) slightly.
In turn, the oblate (rim) now performs much better, as does the prolate (tip).
This clearly indicates that the adhesion dynamics and the actin-ratcheting dynamics
have different preferences concerning curvature.
Finally, the bottom row of Fig.~\ref{fig_uptake_ellips} shows snapshots 
of the initial stage of engulfment for the various shapes and orientations,
cf.~also supplementary movies 3 \& 4.\\

\noindent {\bf Simultaneous phagocytosis of two beads.} 
It is known that cells can engulf more than one particle at the same time.
For several reasons -- among them, the increase in membrane tension, limited resources, as well as
signaling -- one would expect collective effects to occur.
Recently, a spatial resolution limit for phagocytosis was discussed in \cite{Kress_simultaneous},
using holographic optical tweezers to approach two IgG-coated beads
with controlled distances to macrophage cells. 
While signaling is not included in our modeling framework yet 
(except for local actin stimulation $\propto\beta$), 
we can explore the collective effects arising from geometry and volume conservation.

\begin{figure}[t!]
\centering
\includegraphics[width=0.85\linewidth]{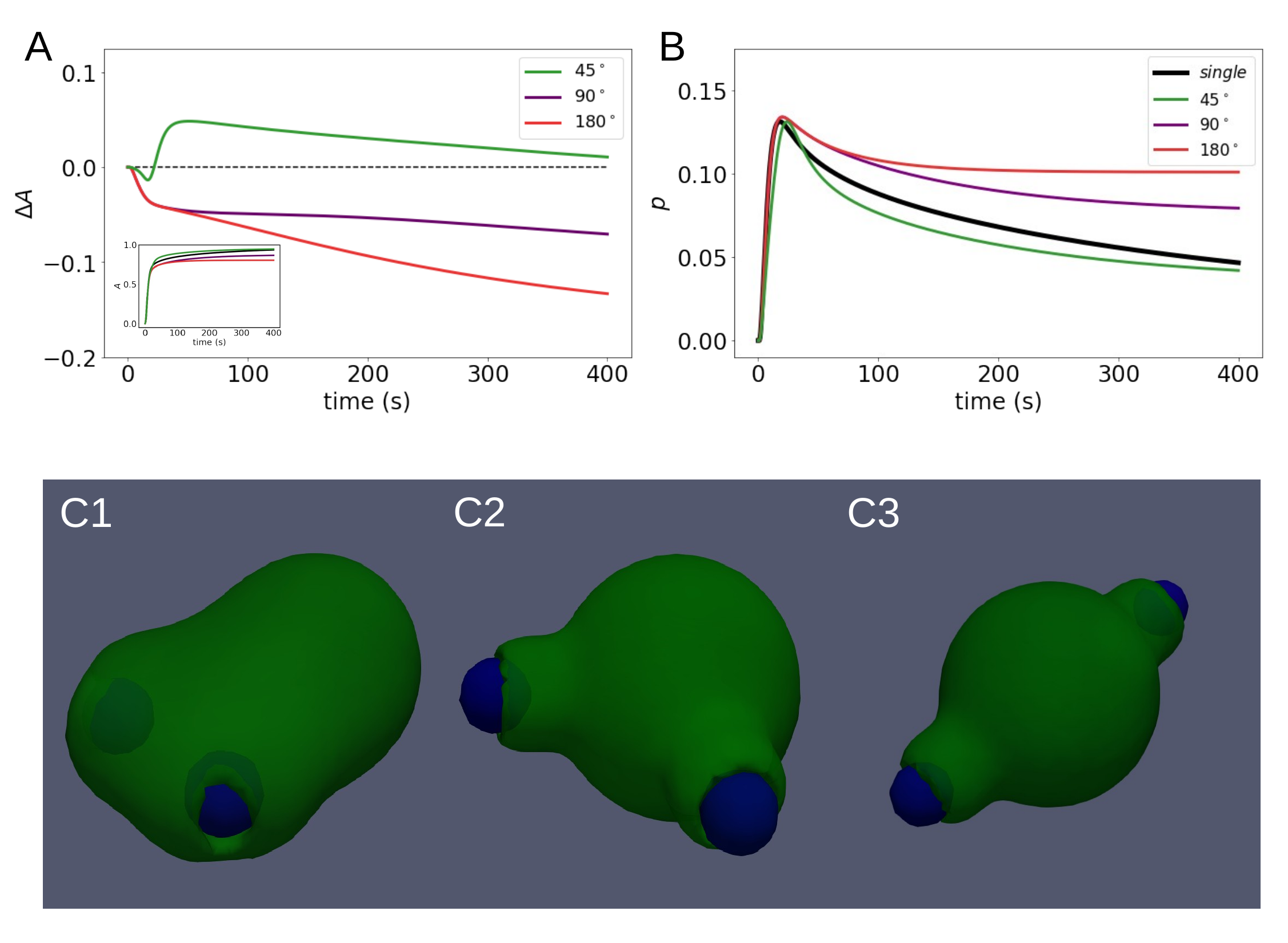}
\caption{\label{fig_uptake_twobeads}
Simultaneous uptake of two spherical beads by an initially spherical cell. 
Investigated are three configurations with angles $45^\circ$, $90^\circ$ and $180^\circ$ 
with respect to the cell's center. 
A) shows the normalized contact area per bead 
relative to the single sphere case 
(the inset shows the absolute relative contact area per bead).
B) shows the normalized overall actin (per particle) as a function of time.  
$45^\circ$ is faster, whereas $90^\circ$ and $180^\circ$ are slower
than single bead uptake. 
C1-C3 show snapshots of the three cases at equal time, $t=50$s.
Parameters: $\alpha=4, \beta=3\alpha$, $r_p=4$, 
rest as given in table \ref{param_table}.
}
\end{figure}

Fig.~\ref{fig_uptake_twobeads} shows results for the simultaneous uptake of 
two spherical beads ($r_p=4$) by an initially spherical cell.
The beads were arranged such that they touch the cell at an
angle of $45^\circ$, $90^\circ$ and $180^\circ$ with respect to the cell's center. 
Panel A shows the difference in contact area with respect to the uptake of
a single sphere, $\Delta A(t)=A_{two}(t)/2-A_{single}(t)$.
Interestingly, the $45^\circ$ positioning (green curve) leads -- after an initial small slowdown --
to faster uptake than in the single bead case, 
before the difference decreases for longer times as $A\rightarrow1$. 
In contrast, the other two configurations lead to slower uptake and to less
overall engulfment (blue and red).
Panel B shows the actin observable $P(t)$, measured per bead, 
showing the same trend: the $45^\circ$ configuration initially needs more time 
but than is faster than for single bead,
while the other two settings get ``stuck'' at higher level of $P$, reflecting 
incomplete engulfment.  Panels C1-C3 show snapshots for the three configurations at equal times,
$t=50$s.\\

\noindent {\bf Phagocytosis by cells spreading on a substrate.} 
In dedicated experiments, phagocytosis can be probed using a freely floating cell
sucked into one pipette while presenting the particle with another pipette
\cite{Herant2006}.
In practice, however, most experiments study cells that already spread on a substrate
when phagocytosis is initiated. 
This means that the initial symmetry of the cell is already broken.
We also studied this situation within our proposed framework, 
cf.~Fig.~\ref{fig_teaser}E. 
The cell was allowed to spread on a substrate (yellow, modeled by another static phase field). 
Then a spherical particle was presented from above and engulfment started.

It should be noted that, in general, both receptor-ligand interaction 
and actin nucleation and polymerization dynamics
are different for the cell-particle ($\kappa, \beta$)
and cell-substrate ($\kappa_s, \beta_s$) contacts.
Due to the use of different phase fields for the particle and the substrate,
this can be easily implemented: 
Fig.~\ref{fig_teaser}E shows the case of a cell 
($r_c=14, \alpha=2$)
that is moderately spreading on the substrate ($\kappa_s=5$, $\beta_s=6$)
and  has a strong interaction with the particle ($\kappa=18$, $\beta=12$) of size $r_p=4$.

We found (data not shown) that the uptake is slower and more incomplete
compared to an initially spherical cell,
since the spreading already increased the cell's surface area,
leaving less freedom to engulf particles. 
We also studied the simultaneous uptake of two spherical beads, 
presented to the spreading cell in a plane parallel to the substrate,
and found the same trends as discussed in the previous section, see supplementary movie 5,
i.e.~accelerated uptake for close-by beads ($45^\circ$) and slowed down uptake for larger angles.

\section{Reduced description of uptake dynamics from perturbation theory}
\label{pert_theory}

One advantage of having a continuum PDE model at hand 
is that it is amenable to perturbation theory to derive reduced models.
These typically have a restricted validity range, but are easier to solve numerically 
and yield additional analytical insight. 
For instance, from a full phase field framework, 
reduced equations for 
membrane waves \cite{reeves2018rotating,Mohammad21},
and cell spreading \cite{Mohammad22} have been derived in 2D and 3D, respectively.

  \begin{figure}
  \centering
  \includegraphics[width=0.99\linewidth]{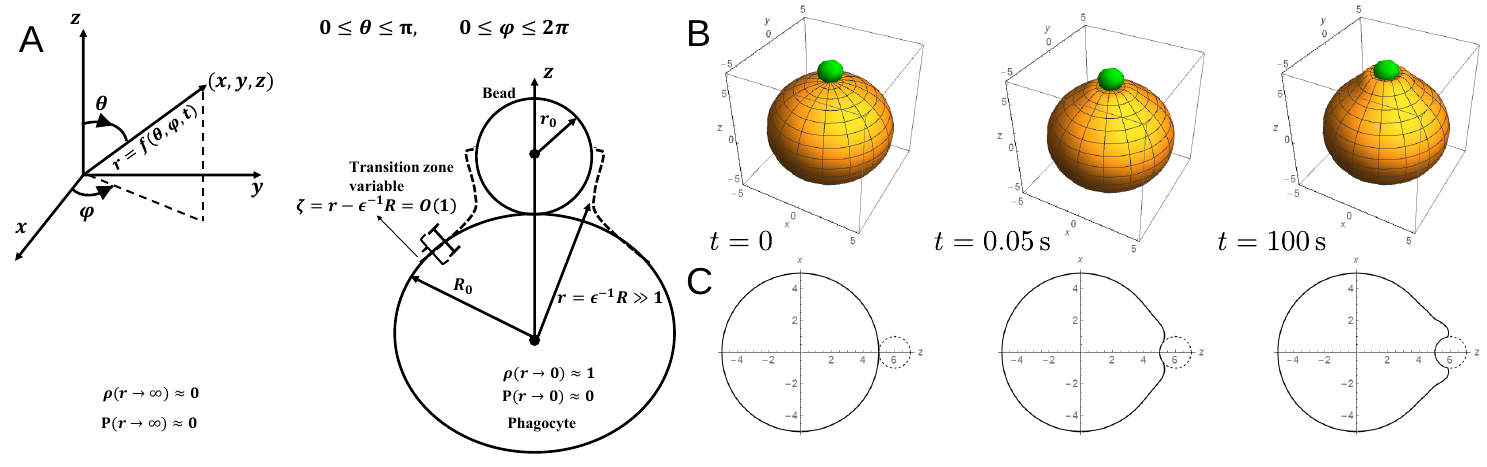}
  \caption{ \label{phag5}
A) A schematic description of the geometry 
specifying the used spherical coordinates,
the transition zone variable and the boundary conditions. 
B) Numerical solution of the reduced model, Eq.~(\ref{eveqmain}),
parameterizing  the phagocyte shape as $R(\ta,t)$  
for times as indicated (initial condition, very early pedestal phase, maximum engulfment). 
The spherical particle (radius $r_0 = 1$) is shown in green and
the gridded surface represents the cell of initial radius $R_0 = 5$.
C) Shown are cross-sections corresponding to the stages of engulfment 
shown in B.
Parameters (see also appendix \ref{perturb_details}): 
$\eps=0.1$; $\beta=25$, $\tilde\alpha=50$ (implying $\alpha=5$),
$\tilde\mu=8$; others as in table \ref{param_table}.
} 
\end{figure}

The geometry considered is the one discussed in section \ref{sphere_sphere}, 
i.e.~an initially perfectly spherical cell touching and starting to engulf a spherical bead,
see Fig.~\ref{phag5}A for a sketch of the geometry and definition of the variables. 
The approach is detailed in \ref{perturb_details}.
In brief, a perturbation method is applied in the sharp interface limit \cite{Mohammad20}:
one assumes that $\epsilon$, the ratio of the cell's phase field interface thickness
to the cell radius, is small, which is typically well fulfilled. 
The model equations (\ref{eq1}) and (\ref{eq2}) had to be slightly modified concerning
the definitions  of $\Phi$ and $\Psi$, to avoid the perturbation to become singular.
Assuming a suitable scaling of the parameters, one can solve the problem
in leading order in $\epsilon$.
From the solvability condition in the next order 
one then obtains a closed evolution equation 
for the cell shape, which 
due to the cylindrical symmetry can be parameterized
by a function  $R(\ta,t)$, 
obeying the dynamics
\begin{eqnarray}\label{eveqmain}
 a \Lambda \partial_tR = -2a D_\rho \mathcal{H} - \tilde{V} + \mathcal{P} - \mathcal{K}.
\end{eqnarray}
Here $ a = (2 D_\rho)^{-1/2}$ and
$\mathcal{H} =\frac{1}{2} \nabla\cdot \hat{\mathbf{n}} = \frac{1}{2} \nabla\cdot \left( \frac{\nabla(r-R)}{|\nabla(r-R)|} \right)$
 is the local mean curvature of the surface $r= R(\ta,t)$.
 The term $  \tilde{V} (t)= \tilde{\mu} \left[ \frac{1}{2} \int_{0}^{\pi} R^3 (\ta,t) \sin\ta  \di \ta - V_0 \right]$
is related to the volume constraint, with $\tilde{\mu}$ the rescaled version of $\mu$.

The third term on the r.h.s. of Eq.~(\ref{eveqmain}), $\mathcal{P}$, 
describes the pushing of actin and can be given explicitly
\begin{equation}
\mathcal{P}(R,\ta,t) = 6\beta \tilde{\alpha}  \Lambda^2 \left(\lambda_p + \frac{\partial_\ta R}{R}\lambda_q  \right)
\int_{-\infty}^{\infty} p_0(\xi) \bar{\rho}_{0}^2(\xi) \di \xi\,.
\end{equation}
It involves the leading order radial phase field profile 
$\bar\rho_0(\xi) = \frac{1}{2} \left[ 1-\tanh\left(\frac{\xi}{\sqrt{8D_\rho}}\right) \right]$
 and the leading order radial actin polarization field
 \begin{equation}
 \label{pfunc}
 p_0(\xi ) =\frac{1}{8}\sqrt{\frac{\tau}{2 D_\rho D_p}} \int_{-\infty}^{\infty} \ex^{-|s|/\sqrt{\tau D_p}} \cosh^{-2} \left( \frac{s - \xi}{\sqrt{8D_\rho}} \right) \di s. 
 \end{equation}
Again, $\tilde{\alpha}$ is a rescaled version of $\alpha$, 
hence $\mathcal{P}\propto\beta\alpha$ is proportional
to both the actin nucleation rate, $\beta$, and the actin ratcheting velocity, $\alpha$.
 
 Finally, the last term in Eq.~(\ref{eveqmain}) describes the effect of 
 ligand-receptor adhesion of the cell on the particle,
 \begin{equation}
   \mathcal{K}= \frac{2a\kappa\Lambda}{D_\rho} G  \exp(-G^2  /D_\rho) \left( \partial_R G - \frac{\partial_\ta G \partial_\ta R}{R^2} \right),
 \end{equation}
 which is proportional to the adhesion strength $\kappa$ and
 where  $G=G(R,\theta)$ is a function 
 capturing the  geometry of the engulfed particle
 i.e.~the effects of the bead phase field $\Phi$.
More details can be found in \ref{perturb_details} and in \cite{Mohammad_arxiv}.

While the specific expressions of the contributions look complicated, Eq.~(\ref{eveqmain}) 
is a single equation on $\theta\in[0,2\pi]$, which is a substantial reduction
compared to two fully 3D equations.
Importantly, this equation makes all relevant contributions nicely transparent:
The first term is proportional to $\sqrt{D_\rho}$, which is the surface tension.
Both membrane deformation 
and the volume constraint impede phagocytosis. 
In turn, actin pushing
-- given by integration of the function in Eq.~(\ref{pfunc}) which is peaked at the interface
where actin is stimulated by the particle -- and adhesion drive phagocytosis.  

Using initial and boundary conditions that guarantee a regular and smooth solution,
\begin{equation}\label{BC_ODE}
  R(t=0)=R_0=R(\ta=0)\,,\,\,R_\ta(\ta=0)=0=R_\ta (\ta=\pi),
\end{equation}
Eq.~(\ref{eveqmain}) can be solved numerically as shown in Fig.~\ref{phag5}B,C. 
Note that this takes just few seconds using Wolfram Mathematica,
compared to several hours running on a GPU for the full 3D model. 

It should be noted, however, that the asymptotic model can not capture a  
complete bead internalization 
and is hence limited to the initial stage of phagocytosis (pedestal phase, cup phase and
early wrapping). This is for several technical reasons: 
first, we assumed a single-valued interface function
(cf.~Eq.~(\ref{ex-rho}), where we wrote explicitly $r=f(\ta,t)$; an implicit parameterization
could be used instead in the future). 
And second, unlike in the full model, we had to include  $\eps$ in the Gaussians
defining the static fields $\Phi$, $\Psi$, cf.~Eqs.~(\ref{Phi}), (\ref{Psi}),
in order to avoid boundary layer problem complications both in time and space.

\section{Conclusion and outlook}

We proposed a three-dimensional phase field framework 
to study the physical part of phagocytosis dynamics.
The processes included so far are (receptor-mediated)
cell-particle adhesion, inducing actin stimulation by the presence of the bead,
which in turn leads to tangential actin polymerization.
We have shown that the framework is very versatile and can study geometries/situations
that have not been studied yet with other approaches,
e.g.~phagocytosis of multiple particles and by spreading cells.
We also found interesting results, for instance, the study of spheroidal particles
revealed that the adhesion dynamics and the actin-ratcheting dynamics
have different preferences concerning curvature.
Finally, we used a perturbation approach in the thin interface limit
to derive a reduced equation for the cell shape $R(\ta,t)$, 
making all important contributions transparent:
membrane deformation 
and volume conservation counteract phagocytosis 
while actin nucleation and pushing, as well as receptor-ligand adhesion, drive it. 

Needless to say that the model still misses most of
the biological components and is hence far from being able to
describe phagocytosis beyond this simple, physical picture. 
Nevertheless, as phase field methods have been applied
to many different systems already, a large toolbox exists to include
relevant processes in the near future, which we like to overview in 
the following.

Membrane: From the physical side, one should definitely 
improve the description of the membrane.
Due to the high curvatures at the advancing phagocytic tip, 
bending rigidity constitutes an important contribution counteracting phagocytosis.
Even more relevant is to include the fact that at the beginning of phagocyotsis, 
membrane tension is reduced due to the opening of membrane reservoirs 
and, much slower, lipid synthesis \cite{Hallett_ironing,Waterman_phago}.
Both explicit surface tension and membrane bending can be included 
into the phase field approach \cite{VoigtJRSI,benjPhysD},
but membrane reservoirs are still hard to model, also in other approaches. 

Cytoskeleton: In the approach proposed so far, 
larger beads stimulate more actin and consequently the uptake 
is not very sensitive to particle size.
While cells can indeed engulf particles of their own size, 
the actin dynamics is obviously oversimplified.
One effect that may be relevant is limited actin resources 
and its time-delayed transport into the phagocytic cup,
amenable to modeling via reaction-diffusion-type additional equations. 
The closure of phagosomes also needs to account for additional
components: since long there is evidence that a
contractile activity closes phagosomes \cite{Swansonmyosin} 
in macrophages, which was related to myosin motors;
see also Ref.~\cite{Gauthier} for a recent review.
While motors apparently are not crucial always  \cite{Waterman_phago},
it would be interesting to include them into the framework,
which is possible by replacing the actin dynamics, Eq.~(\ref{eq2})
by a more detailed ``active gel'' model \cite{JFactivegel}.
Related to this, the advancing phagocytic front also exerts measurable 
forces on the bead \cite{TFM_phago}, and 
elastic effects have been evidenced to 
constitute a bottleneck for phagocytosis advancement \cite{Irmscher_Kress}.
It hence could become relevant to also include elasticity
in the phase field approach, as developed recently \cite{RobertEPJE}.

Signaling: From the biological side, signaling plays an essential role 
in phagocytosis \cite{portals_of_entryNature,EndresRPP,Kress_simultaneous}. 
In fact, phagocytosis is highly concerted and regulated, involves the recruitment of many constituents
and relies on longer-ranged signaling. For instance,
the experimentally observed jump between a receptor-diffusion and an imperfect 
(either receptor- or actin-related) drift regime of the engulfment dynamics
could only be modeled by including a dynamic signaling molecule \cite{Endres_BP14}. 
Signaling processes -- modeled by reaction diffusion-like components -- can be easily
included in the phase-field approach \cite{LevineRappel03,voigtcell},
similar as done recently in the complementary process of 
macropinocytosis \cite{phasefield_macropino} implementing an activator-inhibitor system.

In conclusion, we hope that the here-presented approach can help paving the way
for developing a versatile and more complete physics-based framework 
to model and shed new light on the complex problem of phagocytosis.

\bigskip

\noindent{\bf Data availability statement}\\
\noindent Data that support the findings of this study, as well as the  
code that created the data, 
are available upon reasonable request to the corresponding author.



\appendix

\section{Numerics}
\label{numerics}

The model equations (\ref{eq1}) and (\ref{eq2}) were solved in 3D
with the operator split Fourier pseudo-spectral method on a single GPU using Cuda.
The typical spatial discretization used was $N=256^3$ (implying $\Delta x=0.4\mu$m) 
and a time step of $\Delta t=10^{-3}$s.
As typical model parameters we used the ones given in table \ref{tab:ModelParameter},
except if specified otherwise.

Before introducing the dynamic cell, the static fields representing the particle 
(and, if applicable, the substrate) 
were initialized as described previously \cite{winkler_cell3D}: 
Starting with pixel-wise step functions forming the geometry, 
the field was relaxed for  $t=2$s using a simple phase field dynamics with fixed volume like 
$\partial_t \Phi=D_{\Phi} \Delta\Phi - \Phi(1-\Phi)(\delta[\Phi]-\Phi)$.
This smoothened field was then kept fixed 
as the environment for the introduced cell.
The phase field for the cell was initialized as a spherical tanh-profile of appropriate size.

\begin{table}[ht]
\centering
\begin{tabular}{ l | l   }
\textbf{model parameter} & \textbf{value/range}  \\
\hline 
$\alpha$ & [0.5 - 10] \\
\hline
$\beta$ & [1.5 - 30] $(\beta=3\alpha)$ \\
\hline
$\tau$ & 10   \\
\hline
$D_{\rho}, D_{\mathbf{p}}, D_{\Phi}(=D_{\Psi}) $ & 0.5, 0.2, 0.5 \\
\hline
$\kappa$ & 12 \\
\hline
$\lambda$ & 10 \\
\hline
$r_c$ & 14 \\
\hline
$r_p$ & [2 - 11] \\
\hline
\end{tabular}
\caption{\label{param_table}
Typical model parameters. Time is scaled in seconds ($\rm s$) and
space in microns ($\mu \rm m$). 
The rescaling, using realistic/estimated experimental actin parameters, 
is explained in Refs.~\cite{ziebert2011model,winkler_cell3D}.
}
\label{tab:ModelParameter}
\end{table}%

\section{Details on the perturbation theory}
\label{perturb_details}

We performed a perturbation expansion in the sharp interface 
limit \cite{Mohammad20},
considering the simplest phagocytosis scenario 
of an initially spherical cell of radius $R_0$ touching 
a motionless rigid spherical bead of radius $r_0=\lambda_r R_0$, 
see Fig.~\ref{phag5} A.
For simplicity, we here present the case where $\lambda=0$ 
and where the term $\Phi^2 \mathbf{p}$ in Eq.~(\ref{eq2}) has been omitted,
since this did not affect the results of the reduced model qualitatively. 
A rigorous and more detailed derivation of the next arguments is available in
\cite{Mohammad_arxiv}.

The basic assumption of the perturbation procedure is that the ratio $\eps$ 
of the thickness of the cell's interface 
(i.e., the width of the transition zone, where $\rho$ changes from nearly $1$ to $0$) 
to the characteristic size of the cell (its radius $R_0$) is small,
$\eps\ll 1$, see Fig.~\ref{phag5}A).
For the static phase fields, defining the particle and
the localization of actin polymerization,
we assume
\begin{eqnarray}\label{Phi}
 && \Phi (\textbf{r} ) = \exp  \left[ -\eps^2 \Big( |\textbf{r} - (r_0+R_0)\hat{z}| -r_0 \Big)^2 /D_\rho \right] , \\
 \label{Psi}
 && \Psi (\textbf{r})= \exp  \left[ -\eps^2 \Big( |\textbf{r} - (r_0+R_0)\hat{z}| -r_0 \Big)^2 /(\tau D_p) \right].
\end{eqnarray}
Both functions are $O(1)$ in the vicinity of the spherical bead's boundary 
and  decay exponentially otherwise.
Note that including $\eps$ in the exponentials allows to avoid complications with boundary 
layers both in time and space. This is at the cost of a direct comparison
 to the full simulations becoming difficult.

We now use the standard spherical coordinate system, see Fig.~\ref{phag5}A,
and assume cylindrical symmetry, i.e., independence of all fields on the azimuthal angle $\ph$. Hence we can write $\rho=\rho(r,\ta)$, $\textbf{P}(r,\ta,t)=p\hat{r}+q\hat{\ta}$.
The iso-surface of the cell interface is defined as
\begin{equation}\label{ex-rho}
  \rho\Big( r=f(\ta,t),\ta,t \Big)=\frac{1}{2} .
\end{equation}
In order to balance the front dynamics with curvature we impose the following  scaling 
that describes the slow dynamics
of a large cell \cite{Mohammad20},
\begin{equation}\label{scaling}
  \tilde{t}=\epsilon^2 t, \quad f(\ta,t)=\epsilon^{-1} R(\ta,t), \quad \eps\ll1 .
\end{equation}
The transition zone variable is then defined as
\begin{equation}\label{zeta}
\z=r-f(\ta,t)=O(1).
\end{equation}
We approximate the  nonlocality in the volume constraint by
\begin{equation}\label{nonloc}
   \int \rho\,\di^3 r \sim   \frac{2\pi\eps^{-3}}{3} \int_{0}^{\pi} R^3 (\ta,t) \sin\ta\, \di \ta\,.
\end{equation}
In order to facilitate the analysis, we consider the following scaling of the model parameters
\begin{eqnarray}\label{scal1}
 &&\alpha = \eps \tilde\alpha, \quad   \frac{4\pi\mu}{3}\eps^{-3} = \eps \tilde\mu, \quad 
 \kappa=O(1), \quad \lambda_r=O(1) .
\end{eqnarray}

Now, we introduce expansions of the phase field and the actin field like
 \begin{equation}\label{expan}
  \rho = \rho_0 + \eps  \rho_1+..., \quad p = p_0 + \eps p_1 +...
\end{equation}
and define two auxiliary functions, 
\begin{eqnarray}
\label{Lambda}
\Lambda(\ta,t)&=& 
\left( 1+ \frac{(\partial_\ta R)^2}{R^2}  \right)^{-1/2}\,,\\ 
\label{Phi(z)}
\Phi_p(\tau,D_\rho , D_p, \z  ) 
&=&
\frac{1}{8}\sqrt{\frac{\tau}{2 D_\rho D_p}} \int_{-\infty}^{\infty} \ex^{-|s|/\sqrt{\tau D_p}} \cosh^{-2} \left( \frac{s - \z}{\sqrt{8D_\rho}} \right) \di s,
\end{eqnarray}
 that will appear in the following analysis.

First, we substitute the scalings of length, time 
and the parameters, Eq.~(\ref{scal1})  into Eqs.~(\ref{eq1}), (\ref{eq2}). 
Then we write this system using the transition zone variable, Eq.~(\ref{zeta}), 
using chain rule when needed.  Finally, we apply the asymptotic expansion, 
Eq.~(\ref{expan}) and collect terms of the same order of $\eps$.

The solutions to the leading order are
\begin{eqnarray}\label{}
&& \rho_0 (\z) = \frac{1}{2} \left[ 1-\tanh\left(\frac{\Lambda\z}{\sqrt{8D_u}}\right) \right],\\
&& p_0 (\z) =   \beta \lambda_p  \Lambda\Phi_p(\Lambda\z), \quad q_0 (\z) =  -\beta \lambda_q \Lambda\Phi_p(\Lambda\z)
\end{eqnarray}
for the phase field and the actin polarization, respectively.
Exact expressions for $\lambda_p,$ and $\lambda_q$ are available in 
\cite{Mohammad_arxiv}.

At the next order, $O(\epsilon)$, we derive the equation for $\rho_1(\z)$. 
From its solvability condition, we then get a closed evolution equation 
for the phagocyting cell's interface, $R(\ta,t)$, as given already in Eq.~(\ref{eveqmain}),
\begin{eqnarray}\label{eveq}
&& a \Lambda \partial_t R = -2a D_\rho \mathcal{H} - \tilde{V} + \mathcal{P} - \mathcal{K},
\end{eqnarray}
and discussed in the main manuscript.

\bigskip
\bigskip
\bigskip

\bibliographystyle{unsrt.bst}
\bibliography{cells_ref_3d}

\end{document}